# Electric dipole and quadrupole contributions to valence electron binding in a charge-screening environment

A. D. Alhaidari[(a)] and H. Bahlouli[(b)]

[(a)] *Saudi Center for Theoretical Physics, P. O. Box 32741, Jeddah 21438, Saudi Arabia*
[(b)] *Physics Department, King Fahd University of Petroleum & Minerals, Dhahran 31261, Saudi Arabia*

**Abstract**: We make a multipole expansion of the atomic/molecular electrostatic charge distribution as seen by the valence electron up to the quadrupole term. The Tridiagonal Representation Approach (TRA) is used to obtain an exact bound state solution associated with an effective quadrupole moment and assuming that the electron-molecule interaction is screened by suborbital electrons. We show that the number of states available for binding the electron is finite forcing an energy jump in its transition to the continuum that could be detected experimentally in some favorable settings. We expect that our solution gives an alternative, viable and simple description of the binding of valence electron(s) in atoms/molecules with electric dipole and quadrupole moments. We also ascertain that our model implies that a pure quadrupole-bound anion cannot exist in such a charge-screening environment.

## 1. Introduction

The interaction of a charged point particle (electron) with extended objects such as large molecules or ions is a problem of fundamental interest that received a lot of attention since the early days of nuclear and molecular physics [1]. This phenomenon was modeled by the interaction of the electron as a point charge with the multipole expansion of the molecular electrostatic potential and may result in its capture in an anion bound state whose electronic orbital is of a diffused nature. The interaction is dominated by the weak long-range charge multipole attractive potential that scales as a power law of the inverse distance to the center of the molecule. If the dominant potential is due to the finite dipole moment of the molecule then such a bound state is called dipole-bound state and played an important role in electron-molecule interactions and gave rise to valence-bound anions, which were extensively studied in the literature [1-7]. In recent years the authors obtained a closed form solution of the time-independent Schrodinger equation for an electron in the field of molecule treated as a point electric dipole and deduced the critical value(s) of the dipole strength above which binding could occur for both the ground and excited states [8].

If the molecular dipolar moment vanishes or is negligible while the quadrupole moment is dominant, then it is natural to examine the possibility of electron binding by neutral molecules with significant quadrupole moments to form quadrupole-bound anions in which the long-range charge-quadrupole attractive potential dominates [9-12]. The quadrupole potential varies as $1/r^3$ and is singular for any attractive value of the quadrupole moment. Due to the non-central nature of the potential, which does not allow factorization of the electronic wavefunction, exact closed form solutions were not possible and only asymptotic solutions were known analytically.

One of the first theoretical studies of potentially quadrupole-bound anions was performed by Jordan and Liebman [13]. They considered attachment of an extra electron to $(BeO)_2$ dimer to form the rhombic $(BeO)_2^-$ cluster, which may be considered as a quadrupole-bound anion. This and similar clusters like $(MgO)_2^-$ [14] and $(KCl)_2^-$ [15] were shown to have relatively high electron binding energies and should probably be considered as potential candidates for quadrupole valence-bound anions [9]. However, up to present it has been an experimental challenge to find systems where the excess electron is bound only due to the electric quadrupole moment [10,16] which resulted in a controversy in the literature. The first experimental claim of a quadrupole-bound anion has been obtained in 2004 [17] for the trans-succinonitrile (NC-CH2-CH2-CN) molecule. The observation of an excited quadrupole-bound state for cryogenically cooled cyanophenoxide anions has also been reported recently but it was recognized that the bound state was very weak [18]. Recently, state-of-the-art ab initio calculations on several quadrupole-bound anion candidates [19] concluded that the quadrupole is much weaker than dipole binding as the electron-molecule potential is not dominated by a single component. Coupled cluster ab-initio methods were also used recently [20] to study electron attachment to covalent molecules and ion clusters with vanishing dipole but large quadrupole moments. They found that the quadrupole of the neutral molecular is too weak to bind an excess electron and that there is clearly no such thing as a "critical quadrupole moment".

In this paper, we make a theoretical investigation of the binding of a valence electron in an atomic/molecular system where the charge distribution, as seen by the valence electron, includes contributions coming from the electric monopole (the net charge), electric dipole, and an (effective) electric quadrupole terms. The valence electron interaction with the atom/molecule is assumed to be screened by the outer orbital electrons. Moreover, if the net charge (monopole term in the multipole expansion) is zero, then the solution, if it exists, represents the binding of an electron to a neutral molecule with electric dipole and quadrupole (i.e., an anion). However, our model is not solvable for zero net charge. Nonetheless, taking the zero-charge limit results in a zero energy bound state and hence shows that a pure quadrupole-bound anion is not allowed in the charge-screening environment of our model.

## 2. Formulation of the problem

In the atomic units $\hbar = m = 1$, the time-independent three-dimensional Schrödinger equation for an electron under the influence of an electrostatic potential $V(\vec{r})$ associated with a given charge distribution reads as follows

$$\left[-\tfrac{1}{2}\vec{\nabla}^2 + V(\vec{r}) - E\right]|\psi(\vec{r})\rangle = 0, \qquad (1)$$

where $\vec{\nabla}^2$ is the three-dimensional Laplacian, $E$ is the electron energy and $\psi(\vec{r})$ its wave function. In spherical coordinates, $\vec{r} = \{r, \theta, \varphi\}$, this wave equation becomes

$$\left\{\frac{1}{r^2}\frac{\partial}{\partial r}r^2\frac{\partial}{\partial r}+\frac{1}{r^2}\left[(1-y^2)\frac{\partial^2}{\partial y^2}-2y\frac{\partial}{\partial y}+\frac{1}{1-y^2}\frac{\partial^2}{\partial \varphi^2}\right]-2V(\vec{r})+2E\right\}|\psi(\vec{r})\rangle=0, \quad (2)$$

where $y=\cos\theta$. This equation is separable for potentials with the following general form

$$V(\vec{r})=V_r(r)+\frac{1}{r^2}\left[V_\theta(y)+\frac{1}{1-y^2}V_\varphi(\varphi)\right]. \quad (3)$$

If we write the wavefunction as $\psi(\vec{r})=r^{-1}R(r)\Theta(\theta)\Phi(\varphi)$, then the wave equation (2) with the potential (3) becomes separated in all three coordinates as follows

$$\left[\frac{1}{2}\frac{d^2}{d\varphi^2}-V_\varphi(\varphi)+E_\varphi\right]|\Phi\rangle=0, \quad (4a)$$

$$\left[\frac{1}{2}(1-y^2)\frac{d^2}{dy^2}-y\frac{d}{dy}-\frac{E_\varphi}{1-y^2}-V_\theta(y)+E_\theta\right]|\Theta\rangle=0, \quad (4b)$$

$$\left[\frac{1}{2}\frac{d^2}{dr^2}-\frac{E_\theta}{r^2}-V_r(r)+E\right]|R\rangle=0, \quad (4c)$$

where $E_\varphi$ and $E_\theta$ are the dimensionless angular separation constants. The bound states wavefunction components satisfy the physical boundary conditions: $R(0)=R(\infty)=0$, $\Phi(\varphi+2\pi)=\Phi(\varphi)$ and $\Theta(0)$, $\Theta(\pi)$ being finite. In References [8] and [21], we obtained exact solutions for equations (4) with the potential components $V_\varphi(\varphi)=0$, $V_\theta(y)=Q_d\, y$ and $V_r(r)=-Q/r$ making the total potential function (3)

$$V(\vec{r})=-\frac{Q}{r}+Q_d\frac{\cos\theta}{r^2}, \quad (5)$$

which is that of an electron interacting with an electric dipole of moment $Q_d$ lined up along the positive z-axis and with a net charge $Qe$. We took (and will continue to take) the Bohr radius, $a_0=4\pi\varepsilon_0\hbar^2/me^2=4\pi\varepsilon_0/e^2$, as the unit of length with $m$ and $-e$ being the mass and charge of the electron. For $Q=0$, the solution obtained in [8] and [21] refers to an electron bound to a neutral molecule with a permanent electric dipole moment $Q_d$ (i.e., the dipole-bound anion). In the present study, we extend that work by including higher order contributions coming from the electric quadrupole. Therefore, we consider an electrostatic potential due to the distribution of the atomic/molecular charges and in the multipole expansion of the potential, we include terms up to the linear electric quadrupole where we obtain [22]

$$V(\vec{r})=-\frac{Q}{r}+Q_d\frac{\cos\theta}{r^2}+p\frac{\frac{1}{2}(3\cos^2\theta-1)}{r^3}, \quad (6)$$

where $p$ is the electric quadrupole moment. It is obvious that the quadrupole term destroys separability of the wave equation and makes the search for an exact solution to this problem a highly non-trivial task. In fact, such exact solution is found nowhere in the published literature. Nonetheless, we make two simplifications to the problem so that we end up with an attainable solution that is still interesting and physically useful. The first simplification is to consider an effective electric quadrupole interaction where the angular factor $\frac{1}{2}(3\cos^2\theta - 1)$ is averaged over angular space as $\frac{1}{2}(3<\cos^2\theta>-1) = \eta$, where $\eta$ is a real dimensionless parameter such that $-\frac{1}{2} \leq \eta \leq +1$ [†]. This results in an effective quadrupole potential $Q_q/r^3$, where the effective electric quadrupole moment is $Q_q = \eta p$ whose value could be positive or negative. Therefore, the effective potential (6) regains separability where the angular components of the potential remain as $V_\varphi(\varphi) = 0$ and $V_\theta(y) = Q_d y$ whereas the radial component becomes $V_r(r) = -Q\, r^{-1} + Q_q r^{-3}$. Unfortunately, this simplification is not enough to ease our task of finding an exact solution to the effective problem due to the combination of a highly singular potential in addition to a long-range behavior. Nonetheless, we will show below that by using the TRA we do succeed in obtaining an exact solution if we simplify this model further by making the physically plausible assumption that the electric interaction of the valence electron with the molecule is screened by the outer sub-orbital electron. The size of the screening is difficult to estimate, however, we can assume that the effective nuclear charge has been reduced. Therefore, the effective net charge distribution as seen by the valence electron includes the charge of the nucleus $Ze$ and the suborbital inner electrons that are not contributing to the screening, $-(Z-Q)e$, where $1 \leq Q \leq Z$. Moreover, the screening is taken in the conventional way resulting in a short-range behavior of the interaction with the valence electron that diminishes exponentially as $e^{-\alpha r}$. Consequently, our model has three free parameters; (i) the screening parameter $\alpha$, (iii) the quadrupole parameter $\eta$, and (iii) the screening charge number $Q = 1, 2, 3, .., Z$. These three parameters are to be tuned and fixed by physical observations for a given atom/molecule. We need to mention that we are treating our molecular charge distribution as static neglecting its rotational degrees of freedom within the adiabatic limit where its moment of inertia is considered to be infinite.

The exact solution of the angular equations (4a) and (4b) with $V_\varphi(\varphi) = 0$ and $V_\theta(y) = Q_d y$ subject to the physical boundary conditions were obtained in Refs. [8,21]. Interested readers may consult the cited work and references therein. Here, we just state results that are relevant to the solution of the radial equation (4c). In [8,21], we obtain $2E_\varphi = k^2$, where $k$ is the azimuthal quantum number $k = 0, \pm 1, \pm 2, ...$ On the other hand, the separation constant $2E_\theta$ belongs to the set of eigenvalues of the following tridiagonal symmetric matrix

$$T_{n,m} = \left[\left(n+\ell+\tfrac{1}{2}\right)^2 - \tfrac{1}{4}\right]\delta_{n,m} + Q_d\left[\delta_{n,m+1}\sqrt{\tfrac{n(n+2\ell)}{(n+\ell)^2-1/4}} + \delta_{n,m-1}\sqrt{\tfrac{(n+1)(n+2\ell+1)}{(n+\ell+1)^2-1/4}}\right], \quad (7)$$

---

[†] In an ideal environment, we can calculate the angular average $<\cos^2\theta>$ by using the angular wave function $\Theta(y)$, which was obtained in Ref. [21], as $\langle\Theta(y)|\cos^2\theta|\Theta(y)\rangle$.

where $\ell = |k| = 0, 1, 2, \ldots$. Therefore, for a given value of the electric dipole moment $Q_d$, there is an infinite set of values of $E_\theta$ for each quantum number $\ell$. On the other hand, for a given $\ell$, there is a critical value of $Q_d$ below which the electron cannot stay bounded to the neutral molecule ($Q = 0$) to form a dipole-bound anion. That critical dipole moment is the lowest value that makes the determinant of the matrix $T + \frac{1}{4}I$ vanish, where $I$ is the unit matrix. This corresponds to $2E_\theta = -\frac{1}{4}$, which is the critical singularity strength of the inverse square potential $E_\theta/r^2$ beyond which quantum anomalies appear [23,24]. With this set of values of $E_\theta$ for a given $\ell$, we find below the exact solution of the radial wave equation (4c) where the radial potential $V_r(r)$ behaves near the origin as $-Q\,r^{-1} + Q_q r^{-3}$ whereas far away it gets diminished by an electronic screening of the form $e^{-\alpha r}$.

## 3. The TRA radial solution

The exact angular components of the wavefunction, $\Theta(\theta)$ and $\Phi(\varphi)$, associated with the dipole potential $Q_d \cos\theta/r^2$ are already known (see, for example, Ref. [21] and references therein). Hence, to complete the exact solution of the problem, we need to obtain the radial wavefunction $R(r)$. We start by writing the radial Schrödinger equation for the effective radial potential of Eq. (4c), $V_{eff}(r) = \frac{E_\theta}{r^2} + V_r(r)$, as follows:

$$\left[\frac{d^2}{dr^2} - 2V_{eff}(r) + 2E\right]R(r) = 0, \qquad (8)$$

where we have adopted the atomic units $\hbar = m = 1$ and took $e^2 = 4\pi\varepsilon_0$. Next, we make a transformation to a dimensionless coordinate $x(r) = \coth(\lambda r)$, where $r \geq 0$ and $\lambda$ is a real positive scale parameter. Thus, $x \geq 1$ and Eq. (8) is mapped into the following second order differential equation in terms of $x$

$$\lambda^2(1-x^2)\left[(1-x^2)\frac{d^2}{dx^2} - 2x\frac{d}{dx} + \frac{E - V_{eff}(x)}{2\lambda^2(1-x^2)}\right]R(r(x)) = 0, \qquad (9)$$

In Ref. [25], we used the TRA to obtain an exact series solution of this equation. In the TRA, the solution of the wave equation is written as infinite (or finite) series of square integrable functions, which are required to be complete and to produce a tridiagonal matrix representation for the wave operator [26]. Consequently, the matrix wave equation gives rise to a three-term recursion relation for the expansion coefficients of the series. We solve the recursion in terms of orthogonal polynomials whose properties (e.g., weight function, zeros, asymptotics, etc.) give the physical properties of the system such as the bound states energies, the density of states, the scattering phase shift, etc. In Ref. [25], the solution of Eq. (9) was obtained as the series $|R(r)\rangle = \sum_{n=0}^{N} f_n |\phi_n(x)\rangle$, where $\{\phi_n(x)\}$ is the set of square integrable basis with elements

$$\phi_n(x) = c_n (x-1)^{\frac{\mu}{2}} (x+1)^{\frac{\nu}{2}} P_n^{(\mu,\nu)}(x), \tag{10}$$

where $P_n^{(\mu,\nu)}(x)$ is the Jacobi polynomial, $\mu > -1$, $\mu + \nu < -2N - 1$, $n \in \{0,1,2,...,N\}$ and $N$ is a non-negative integer. Therefore, $\nu$ must be negative whereas the normalization constant is taken as $c_n = \sqrt{\frac{\sin \pi(\mu+\nu+1)}{2^{\mu+\nu+1} \sin \pi \nu}} \sqrt{(2n+\mu+\nu+1) \frac{\Gamma(n+1)\Gamma(n+\mu+\nu+1)}{\Gamma(n+\mu+1)\Gamma(n+\nu+1)}}$. The TRA compatible potential in Eq. (8) is obtained in [25] as

$$\frac{2}{\lambda^2} V_{eff}(r) = A[\coth(\lambda r) - 1] - \frac{B}{\sinh^2(\lambda r)} + C \frac{\cosh(\lambda r)}{\sinh^3(\lambda r)}, \tag{11}$$

where $\{A, B, C\}$ are real dimensionless constants such that $A \leq -\frac{1}{2}$ and the basis parameters $\mu$ and $\nu$ are related to the potential parameter $A$ and the energy as follows

$$\mu = \sqrt{-\varepsilon}, \quad \nu = -\sqrt{-\varepsilon - 2A}, \tag{12}$$

where $\varepsilon = 2E/\lambda^2$. Therefore, reality dictates that the energy is negative corresponding to bound states. Now, if we write the expansion coefficients as $f_n = f_0 P_n$, then $P_0 = 1$ and $P_n$ becomes the orthogonal polynomial $H_n^{(\mu,\nu)}(z^{-1}; \omega, \sigma)$ defined in [25,27,28] by its symmetric three-term recursion relation

$$(\cosh \omega) H_n^{(\mu,\nu)}(z^{-1}; \omega, \sigma) = D_{n-1} H_{n-1}^{(\mu,\nu)}(z^{-1}; \omega, \sigma) + D_n H_{n+1}^{(\mu,\nu)}(z^{-1}; \omega, \sigma)$$
$$+ \left\{ z^{-1}(\sinh \omega) \left[ \left(n + \frac{\mu+\nu+1}{2}\right)^2 + \sigma \right] + F_n \right\} H_n^{(\mu,\nu)}(z^{-1}; \omega, \sigma) \tag{13}$$

where $z = -\sqrt{B^2 - C^2}$, $\cosh \omega = B/C$, $\sigma = -\frac{1}{4}$, $F_n = \frac{\nu^2 - \mu^2}{(2n+\mu+\nu)(2n+\mu+\nu+2)}$ and $D_n = \frac{2}{2n+\mu+\nu+2} \sqrt{\frac{(n+1)(n+\mu+1)(n+\nu+1)(n+\mu+\nu+1)}{(2n+\mu+\nu+1)(2n+\mu+\nu+3)}}$. Thus, reality dictates that $B/C \geq 1$, which is in fact what is needed to produce a potential configuration that can support bound states. The overall factor $f_0$ is the square root of the positive definite weight function of $H_n^{(\mu,\nu)}(z^{-1}; \omega, \sigma)$. Unfortunately, the analytic properties of this polynomial are not found in the mathematics literature. This is still an open problem in orthogonal polynomials [28]. Nonetheless, these polynomials could be written explicitly to all degrees, albeit not in a closed form, using the recursion relation (13) and the initial values $H_0^{(\mu,\nu)} = 1$, $H_{-1}^{(\mu,\nu)} \equiv 0$.

It is obvious that the potential (11) is singular at the origin with $r^{-1}$, $r^{-2}$ and $r^{-3}$ singularities of strength $A$, $-B$ and $C$, respectively. However, it has a short range since it decays exponentially away from the origin as $e^{-2\lambda r}$. Thus, $\lambda$ gives a measure of the extent of the range of this potential (i.e., the screening of the electric interaction where the screening parameter $\alpha = 2\lambda$). Near the origin, the potential function (11) behaves as follows

$$V_{eff}(r) \approx \frac{1}{2}\left[\frac{\lambda A}{r} - \frac{B}{r^2} + \frac{C/\lambda}{r^3}\right]. \tag{14}$$

On the other hand, the effective radial potential of Eq. (4c) in our model with a net charge $Q$, electric dipole moment $Q_d$, and effective electric quadrupole moment $Q_q$ is

$$V_{eff}(r) = \frac{E_\theta}{r^2} + V_r(r) = -\frac{Q}{r} + \frac{E_\theta}{r^2} + \frac{Q_q}{r^3}. \tag{15}$$

Comparing (14) and (15) suggests that we can use potential (11) to model the screening of the atomic/molecular charge distribution whose net charge is $Q = -\lambda A/2$, effective electric quadrupole moment is $Q_q = C/2\lambda$ and with an electric dipole moment $Q_d$ whose associated angular contribution is $E_\theta = -B/2$. Consequently, the bound state solution obtained here could be considered as an alternative, viable and simple description of the binding of valence electron(s) in atoms/molecules with electric dipole and quadrupole moments. Another conclusion from our findings goes as follows. The short-range behavior due to screening will permit only a finite number of bound states as will be confirmed by the results of our calculation below. Consequently, transition of the valence electron to the continuum will not be smooth but a jump is expected due to the finite energy difference between the highest excited state and the continuum. In a favorable experimental setting, it might be possible to detect this energy jump. However, in the absence of screening ($\lambda = 0$), the long-range behavior of the interaction allows for an infinite number of bound states with infinitesimally small spacing and densely packed at the top of the energy spectrum near the continuum. This makes transition of the valence electron to the continuum smooth. Moreover, since bound state solution requires $B/C \geq 1$, then this implies that $E_\theta/Q_q \leq -\lambda$ and since $E_\theta$ depends on the azimuthal quantum number $\ell$, then this puts a bound on these quantum numbers ($\ell = 0,1,2,..,\ell_{max}$) that depends on the effective quadrupole moment $Q_q$. Additionally, since $A \leq -\frac{1}{2}$ and $A = -2Q/\lambda$, then for a given charge $Q$ the value of the screening parameter must be chosen from within the range $0 < \lambda \leq 4Q$.

We could, in principle, use our model as an extension to, and/or generalization of, the celebrated dipole-bound anion problem if we could solve it for $Q = 0$. However, as seen above the solution requires that the net charge of the atom/molecule be non-zero and positive (i.e., ionized molecule) since $A \leq -\frac{1}{2}$ and $Q = -\lambda A/2$. Nonetheless, in the limit of extremely week screening (i.e., $\lambda \to 0$) a quadruple-bound anion is, in principle, possible since the electric charge on the molecule tends to zero but we must also let $C \to 0$ such that the ratio $C/\lambda \to 2Q_q$. However, the resulting binding energy, which is $E = \lambda^2 \varepsilon/2$, will tend to zero implying that the electron cannot stay bound to form a stable quadrupole-bound anion. Therefore, we conclude that a pure quadrupole-bound anion cannot exist in such an environment. These findings are consistent with the experimental observations that there has been no firm evidence for the existence of weakly bound anions due to the quadrupolar interactions alone [17].

## 4. Numerical Procedure

In the absence of knowledge of the analytic properties of the orthogonal polynomial $H_n^{(\mu,\nu)}(z^{-1};\omega,\sigma)$, we resort to numerical techniques to obtain the energy spectrum. Due to the energy dependence of the basis elements via the parameters $\mu$ and $\nu$ as shown by Eq. (12), we may use one of two numerical procedure (i) the "*potential parameter spectrum* (PPS)" techniques or (ii) the direct diagonalization of the Hamiltonian matrix (HMD) in an energy independent basis similar to those in (10). In the latter, we use Gauss quadrature integral approximation associated with the Jacobi polynomials. We will not give details of these two techniques. Interested readers may consult Refs. [29,30] for a description of the PPS technique. In the Appendix, we give a brief outline of how to obtain the matrix elements of the Hamiltonian and compute the energy spectrum using Gauss quadrature integral approximation. To develop our model for a given atom/molecule with an atomic/molecular charge number $Z$, an electric dipole moment $Q_d$ and an electric quadrupole moment $p$, we follow a procedure with the following steps:

1. Construct a large enough size Table for the eigenvalues $\{2E_\theta\}$ of the tridiagonal matrix $T$ in (7) with the azimuthal quantum number $\ell = 0,1,2,...$ For each $\ell$ label the sorted eigenvalues $E_\theta$ by an integer $m = 0,1,2,...$ Therefore, a pair of the azimuthal quantum numbers $\ell$ and $m$ uniquely identifies $E_\theta$.
2. Choose a value for the net charge (monopole term) $Q$ from the set $\{1,2,..,Z\}$.
3. Choose a value for the charge screening parameter from within the range $0 < \lambda \leq 4Q$.
4. Choose a value for the electric quadrupole parameter from within the range $-\frac{1}{2} \leq \eta \leq +1$.
5. Impose the constraint $E_\theta/\eta p \leq -\lambda$ on the Table obtained in step 1 above. The result is a sub table for $E_\theta$ whose columns are indexed by $\ell \in \{0,1,2,..,\ell_{max}\}$ with each $\ell^{th}$ column having a finite number of rows $m \in \{0,1,2,..,m_\ell\}$.
6. Take the following potential parameters: $A = -2Q/\lambda$, $B = -2E_\theta$ and $C = 2\lambda\eta p$, where $E_\theta$ is one of the values from the sub table in the previous step labeled by the pair of azimuthal quantum numbers $\ell$ and $m$.
7. With the potential parameters $\{A,B,C\}$ from the previous step, use one of the two methods, PPS or HMD, to compute the binding energy of the valence electron to the atom/molecule and compare to physical observations. Repeat the calculation for all values of $E_\theta$ from the sub table in step 5 above.
8. If an agreement with experimental observations is not reached then go back to step 2 above to tune the parameters and repeat the calculation with an alternate set $\{Q,\lambda,\eta\}$.

9. Once an agreement is reached then register the set of model parameters $\{\lambda,\eta\}$ and quantum numbers $\{Q,\ell,m\}$ that will be used to describe the binding of the valence electron to the atom/molecule whose atomic/molecular charge number $Z$, electric dipole moment $Q_d$ and electric quadrupole moment $p$.

As an example, we consider a hypothetical molecule with charge number $Z=12$, electric dipole moment $Q_d=35$ and electric quadrupole moment $p=7$. Table 1 gives the energy spectrum of the valence electron for a model parameters $\{Q,\lambda,\eta\}=\{8,0.2\,Q,0.3\}$ and for the listed azimuthal quantum numbers $\{\ell,m\}$ corresponding to the shown values of $E_\theta$. The Table shows that favorable settings to measure the energy jump to the continuum are obtained if the valence electron is in a state corresponding to the azimuthal quantum numbers $\{\ell,m\}=$ (0,2), (1,1) and (2,0).

## 5. Conclusion

In this work, we have focused on how to describe the weak binding of an external electron to a stable positively charged molecule using a single electron Schrodinger equation with an effective interaction potential that involves the attractive Coulomb interaction at large distances in addition to higher order dipole and quadrupole of a static molecular charge distribution. We have used the tridiagonal representation approach to find the binding energy of the valence electron to an atom/molecule with an electric dipole and quadrupole but with charge screening coming from suborbital electrons. We showed (as it is always the case for the screened Coulomb interaction) that the number of states available for binding the electron is finite, which might give rise to an energy jump in its transition to the continuum that could be detected experimentally under some favorable conditions. The model we presented is not solvable for zero net charge and hence cannot describe an anion. However, taking the limit of zero net charge in the model results in a vanishing binding energy, which means that our model asserts that a quadrupole-bound anion is not allowed in a screening environment. These findings are consistent with experimental observations that there has been no clear evidence for the existence of weakly bound anions due to the quadrupolar interactions alone [17].

Our present model is mainly influenced by the asymptotic inverse cubic behavior of the quadrupolar molecular potential and neglects details related to the short-range nature of the potential related to the anisotropy of the molecular charge distribution and did not take into consideration the rotational motion of the molecule. Regardless of the above shortcomings, we expect that our solution gives an alternative, viable and simple description of the binding of valence electron(s) to static positively charged atoms/molecules having finite electric dipole and quadrupole moments.

# Appendix: Hamiltonian matrix, Gauss quadrature and evaluation of the energy spectrum

The matrix elements of the Hamiltonian operator in the energy independent basis elements (10) is obtained in Ref. [25] as

$$\langle \phi_m | H | \phi_n \rangle = \lambda \int_0^\infty \phi_m(x) H \phi_n(x) dr$$
$$= \frac{\lambda^2}{2} c_m c_n \int_1^\infty (x-1)^\mu (x+1)^\nu W(x) P_m^{(\mu,\nu)}(x) P_n^{(\mu,\nu)}(x) dx \quad (A1)$$

where $W(x) = \frac{1}{2}\frac{\mu^2}{1-x} + \frac{1}{2}\frac{\nu^2 + 2A}{1+x} - \left(n + \frac{\mu+\nu+1}{2}\right)^2 + \frac{1}{4} - B + Cx$ and we used the integral transform $\lambda \int_0^\infty \ldots dr = \int_1^\infty \ldots \frac{dx}{x^2-1}$. The recursion relation and orthogonality of the Jacobi polynomial show that all terms in $W(x)$ that are proportional to $x^k$, where $k$ is a non-negative integer, will produce exact matrix elements that are $2k+1$ diagonal (i.e., a banded diagonal matrix whose diagonal band is of width $2k+1$). Therefore, only the first two terms in $W(x)$ will produce a matrix with non-zero entries everywhere, which needs to be evaluated numerically. We can eliminate one or both of these terms by choosing $\mu = 0$ and/or $\nu = -\sqrt{-2A}$. However, if we eliminate both then the condition $\mu + \nu < -2N - 1$ will dictate that $N < \frac{1}{2}\left(\sqrt{-2A} - 1\right)$, which restricts severely the size of the matrices and hinder the accuracy of the calculation. Therefore, we eliminate only the first term, which is singular at $x = 1$, and choose $\mu = 0$ while keeping $\nu$ arbitrary but such that $\nu < -2N - 1$.

To calculate the matrix elements that represent the non-zero second term in $W(x)$, we use Gauss quadrature integral approximation associated with the Jacobi polynomial. The details of this method are found in Refs. [31,32]. The essence of the method might be summarized as follows. Let $X_{n,m}$ be the matrix elements obtained by choosing $W(x) = 2x/\lambda^2$ in Eq. (A1). Then by using the recursion relation and orthogonality of the Jacobi polynomial it is easy to show that $X$ reduces to the following tridiagonal symmetric matrix

$$X_{n,m} = F_n \delta_{n,m} + D_n \delta_{n,m-1} + D_{n-1} \delta_{n,m+1}, \quad (A2)$$

where the numbers $\{F_n, D_n\}$ are defined in section 3 below Eq. (13) (with $\mu = 0$). Now, let $\{\tau_n\}_{n=0}^N$ be the eigenvalues of the $(N+1) \times (N+1)$ truncated version of the matrix $X$ and $\{\Lambda_{m,n}\}_{m=0}^N$ be its normalized eigenvector associated with the eigenvalue $\tau_n$. Then an approximate evaluation of the matrix elements that represent any function $W(x)$ is

$$W_{n,m} \simeq \frac{\lambda^2}{2} \left(\Lambda \cdot \Omega \cdot \Lambda^T\right)_{n,m}, \quad (A3)$$

where $\Omega$ is a diagonal matrix with elements $\Omega_{n,m} = W(\tau_n)\delta_{n,m}$. Therefore, the evaluated Hamiltonian matrix has finally the following elements

$$\frac{2}{\lambda^2}H_{n,m} \simeq -\frac{1}{4}\left[(2n+\nu+1)^2 + 4B - 1\right]\delta_{n,m} + CX_{n,m} + \frac{\nu^2 + 2A}{2}\left(\Lambda \cdot \Omega \cdot \Lambda^{\mathrm{T}}\right)_{n,m}, \quad (A4)$$

where $\Omega_{n,m} = \frac{\delta_{n,m}}{1+\tau_n}$. The size of this matrix (number of basis elements) is $N+1$ with $n,m \in \{0,1,2,...,N\}$ and $\nu < -2N-1$. To calculate the energy spectrum, we write the energy eigenvalue equation (wave equation) $H|\psi\rangle = E|\psi\rangle$ in the energy independent basis as $\sum_m f_m H|\phi_m\rangle = E\sum_m f_m|\phi_m\rangle$. Projecting from left by $\langle\phi_n|$, we obtain $\sum_m H_{n,m} f_m = E\sum_m \omega_{n,m} f_m$, where $\omega$ is the basis overlap matrix whose elements are $\omega_{n,m} = \langle\phi_n|\phi_m\rangle$. Consequently, the energy spectrum is calculated as the generalized eigenvalues of the matrix equation $H|f\rangle = E(\omega|f\rangle)$. To obtain the overlap matrix $\omega$, we follow the same Gauss quadrature procedure outlined above giving

$$\omega_{n,m} \simeq -\left(\Lambda \cdot \tilde{\Omega} \cdot \Lambda^{\mathrm{T}}\right)_{n,m}, \quad (A5)$$

where $\tilde{\Omega}_{n,m} = \frac{\delta_{n,m}}{1-\tau_n^2}$. Finally, we vary the value of $\nu$ until a plateau of computational stability of the energy spectrum is reached for large enough $N$. It turns out that the plateau of computational stability (range of values of $\nu$ with no significant change in the result) is larger for lower bound states.

## Table Caption

**Table 1**: The energy spectrum of the valence electron (in units of $-\frac{1}{2}\lambda^2$) for a hypothetical molecule with $\{Z, Q_d, p\} = \{12, 35, 7\}$ in the atomic units. The physical parameters are taken as $\{Q, \lambda, \eta\} = \{8, 0.2\,Q, 0.3\}$ so only $Q-1=7$ outmost suborbital electrons are contributing to the screening. The azimuthal quantum numbers $(\ell, m)$ and the corresponding value of $E_\theta$ are shown and we have used the HMD method outlined in the Appendix with a basis size of 100. For the displayed accuracy, the plateau of computational stability was taken as $\nu \in [-2N-1, -2N-21]$ with $N=100$.

**Table 1**

| $(\ell, m)$ | $E_\theta$ | $\varepsilon$ |
|---|---|---|
| (0,0) | −29.336747 | 458.70220564<br>201.89558358<br>83.02498131<br>29.27760795<br>6.7669578<br>0.175 |
| (0,1) | −18.028503 | 98.32461711<br>30.95016950<br>6.3724845<br>0.099 |
| (0,2) | −7.279835 | 4.41151545 |
| (1,0) | −23.414383 | 225.280359848<br>87.366542332<br>29.136472495<br>6.28034949<br>0.109 |
| (1,1) | −12.362103 | 29.004443567<br>5.05763725 |
| (2,0) | −17.193036 | 84.506325074<br>25.319280377<br>4.49358085 |
| (2,1) | −6.360787 | 2.517451 |
| (3,0) | −10.624897 | 17.45549174<br>1.86655 |
| (4,0) | −3.653598 | 0.074 |